\documentclass[prl,showpacs,twocolumn,aps,superscriptaddress,a4paper]{revtex4-1}
\usepackage{dcolumn,amssymb,amsmath,amsfonts,graphicx,latexsym,color}

\begin{document}

\title{Randomly Fluctuating Potential Controlled Multistable Resonant Tunneling Current through a Quantum Dot}

\author{Pei Wang}
\email{wangpei@zjut.edu.cn}
\affiliation{Institute of Applied Physics, Zhejiang University of Technology, Hangzhou 310023, China}
\date{\today}
\author{Gao Xianlong}
\email{gaoxl@zjnu.edu.cn}
\affiliation{Department of Physics, Zhejiang Normal University, Jinhua 321004, China}
\author{Shaojun Xu}
\affiliation{School of Economics and Management, Zhejiang Sci-Tech University, Hangzhou 310018, China}

\date{\today}

\begin{abstract}
We study the transport through a quantum dot subject to a randomly fluctuating potential, generated by a sequence of pulses in the gate voltage with the help of the autoregressive model. We find that the tunneling current is multistable when the fluctuating potential with a finite correlation time is applied before the non-equilibrium steady state is built up. The non-equilibrium stationary current is heavily dependent on the history of the fluctuating potential during the transient period if the potential has a finite correlation time.  Furthermore, the averaged current over the path of the fluctuating potential is a function of its strength and correlation time. Our work therefore provides a robust theoretical proposal for the controlling of the non-equilibrium stationary current through a quantum dot in a randomly fluctuating potential.
\end{abstract}

\maketitle

{\it Introduction.}-The study of the electronic transport in mesoscopic systems is one of today's most active research areas in condensed matter physics. A typical current-carrying system comprises of two electron reservoirs of different temperatures or chemical potentials, between which the electrons continuously flow~\cite{datta}. Whether the stationary current is uniquely determined by the temperatures and chemical potentials of the reservoirs is a fundamental problem in the quantum transport~\cite{dhar,khosravi,doyon,bokes,wilner13,alexandrov,albrecht}.

One of the simplest models that can carry a non-equilibrium stationary current is the resonant level model, typically describing the coherent transport through a nanostructure~\cite{mehta}, e.g., a quantum dot made from the semiconductor heterostructure. In the wide band limit, the tunneling current $I$ as a function of the voltage bias $V$ is well known to be $I=(\Gamma/\pi) \arctan [V/(2\Gamma)]$ at zero temperature~\cite{jauho94}, independent of the initial conditions, where $\Gamma$ denotes the level broadening. In recent years, the transport through nanostructures subject to time-dependent potentials, called the driven quantum transport, attracted much attention~\cite{grifoni,platero,kohler05,huneke,kwapinski,vogt,hsu,stehlik,sau,souza,hammer}. In these studies, a deterministic driving, especially a time-periodic one, is used, in which the Keldysh formalism, the Floquet approach, or the transfer matrix approach can be applied (see Ref.~\cite{platero} for a review of methods). However, little is known about the non-equilibrium stationary current when the level position of a quantum dot fluctuates randomly in time, which can be caused by the interactions between the electrons in the dot and the phonons in the environment~\cite{german94} or by a manually generated fluctuating potential.

The effect of the electron-phonon interaction on the electron transport has been discussed recently by several authors~\cite{wilner13,alexandrov,albrecht,mitra}, while there is still controversy on the existence of the multistability of the stationary currents. An ab-initio consideration of the electron-phonon interaction is theoretically difficult. An alternate way is to study the dot in a fluctuating potential which emerges as tracing out the environmental degrees of freedom, a way generally employed in studying the dynamics of an open classical system, e.g., the Brownian motion~\cite{coffey}.

A fluctuating level can also be implemented manually by generating a sequence of pulses in the gate voltage controlling the level position. We realize the sequence of pulses with the specified statistics by the autoregressive model~\cite{kaplan}, a representation of a type of random process, which has been frequently discussed in the time-series analysis and widely used in the case when uncertainty dominates, e.g., in nature and in econometrics~\cite{tsay}.

In this letter, we find that the fluctuating potential at the resonant level leads to the multistability of the stationary current which depends on the history of the potential (i.e., the trajectory of the potential in course of time) during the transient period. Our results show that the current through the resonant level model depends not only upon the parameters of the reservoirs, but also upon the way of reaching the non-equilibrium steady state. At the same time, our results provide a new perspective of controlling the tunneling current through a quantum dot, serving as a candidate for the single electron transistor in future integrated circuits~\cite{heinzel}.

{\it Formalism.}-We consider a quantum dot with a fluctuating potential $V_g(t)$ applied to the level, which is coupled to two semi-infinite leads labeled by $L$ and $R$, with the chemical potentials $\mu_L$ and $\mu_R$, respectively. The voltage bias is $V=\mu_L-\mu_R$. The total Hamiltonian is written as
\begin{equation}
 \hat H (t) =  \hat H_L + \hat H_R + \hat H_d(t) + \hat H_V(t),
\end{equation}
where $\hat H_L= -g \sum_{j=-\infty}^{-2} \left( \hat c^\dag_j \hat c_{j+1}+h.c. \right)$ and $\hat H_R=-g\sum_{j=1}^{\infty} \left(\hat c^\dag_j \hat c_{j+1} + h.c.\right)$ describe the left and right leads respectively with $g$ the hopping amplitude and $\hat c_j$ the electron annihilation operator at the site $j$, $\hat H_d(t) = V_g(t) \hat c^\dag_0 \hat c_0 $ the fluctuating level and $\hat H_V(t)=g_c(t) \left( \hat c^\dag_{-1}\hat c_0 + \hat c^\dag_1 \hat c_0 + h.c.\right)$ the coupling between the level and two leads. The coupling $g_c(t)=g_c\theta(t)$ is switched on at the initial time $t=0$. At $t<0$, two leads are at their own equilibrium states, and the potential $V_g(t<0)=\infty$ so that there is no electron at the level. And $V_g=0$ denotes the resonant position, where the level broadening is known to be $\Gamma=2g_c^2/g$~\cite{meisner09}. We take the wide band approximation by setting $g=10g_c$.

At $t=0$, the fluctuating potential is switched on, simulated by the famous autoregressive model~\cite{tsay} of a stochastic process:
\begin{equation}\label{armodel}
 V_g(t+\Delta t) = \phi V_g(t) + R W_t,
\end{equation}
where $0\leq \phi<1$ is a parameter, $W_t$ a white noise with the standard normal distribution and $R$ the strength of the fluctuation. $\Delta t$ denotes the time step and is set to be much smaller than the time scale over which the system changes significantly. The autoregressive model has been intensively studied in the time-series analysis with the average of $V_g(t)$ being zero and the autocovariance being $\langle V_g(t_1) V_g(t_2)\rangle = R^2/(1-\phi^2) exp(-|t_2-t_1|/\tau ) $, where the correlation time is defined as $\tau = -\Delta t/\ln \phi$. The fluctuating potential satisfying Eq.~(\ref{armodel}) is featured by its stationarity and exponentially decaying two-time correlation.

The main quantity we are interested is the tunneling current,
\begin{equation}
 I(t) = -g_c \left( \textbf{Im} \langle \hat c^\dag_{-1}(t) \hat c_0(t) \rangle + \textbf{Im} \langle \hat c^\dag_{0}(t) \hat c_1 (t) \rangle \right),
\end{equation}
which is the averaged current of the left and right leads and the expectation value is done to the initial state. Due to the presence of a time-dependent potential, it is impossible to solve it analytically. We solve it numerically by the excitation operator method~\cite{wang13}, which is accurate and efficient in obtaining both the transient and stationary currents with a time-dependent Hamiltonian (See the details of the method in the supplementary materials).

The current depends on the potential $V_g(t)$, a stochastic time series, in which the noise term $W_t$ is created by a random number generator. For the given parameters $R$ and $\tau$, the potential $V_g(t)$ is generated for statistically significant number of times (more than one thousand times) to study the average and the standard deviation of the current. A stationary current $I_i$ in the $i$-th simulation is calculated for each simulation of $V_g(t)$. The averaged current is obtained as a path ensemble average on the total number of simulations $N$, i.e., $I= \lim_{N\to \infty}(1/N) \sum_{i=1}^N I_i$,
and similarly the standard deviation of current $\sigma_I= \lim_{N\to\infty} (1/N)\sqrt{\sum_{i=1}^N (I_i-I)^2}$.

\begin{figure}
\includegraphics[width=1.0\linewidth]{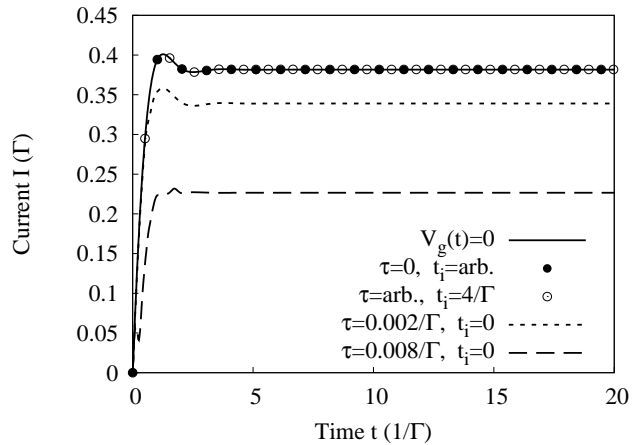}
\caption{The evolution of the current after the coupling between leads and the level is switched on in the presence of different fluctuating potentials, compared with that of no fluctuating potentials (solid line).
We illustrate the case with a potential of a white noise ($\tau=0$) applied in arbitrary (shorted as arb.) time (solid circle), and the one with an arbitrary potential switched on at $t_i = 4/\Gamma$ larger than the relaxation time of currents (empty circle). They both coincide with the $I-t$ curve of $V_g(t)=0$. However, a fluctuating potential with a finite correlation time switched on during $[0, 2/\Gamma]$ significantly suppresses the current. The correlation time is $\tau=0.002/\Gamma$ (short dashed line), and $\tau=0.008/\Gamma$ (long dashed line), respectively. And the voltage bias and the strength of fluctuation are set to $5\Gamma$ in all of the above simulations except for $V_g(t)=0$.}\label{fig:evolution}
\end{figure}
{\it History-dependent stationary current.}-After the coupling between leads and the level is switched on at $t=0$, the current experiences a transient period before relaxing to its stationary value. The turn-on and turn-off times of the fluctuating potential are denoted by $t_i$ and $t_f$ in the following, respectively. We find that applying a fluctuating potential during the transient period will drive the system into a non-equilibrium steady states (NESS) of a suppressed stationary current, distinguished from that of no fluctuating potentials (i.e., $V_g(t\ge 0)\equiv 0$). And the suppressed current survives even after the fluctuating potential is closed after a certain time (see Fig.~\ref{fig:evolution}). 

The suppressed stationary current depends on two critical conditions. First, the fluctuating potential must be applied before the system relaxes to its steady state, i.e., $t_i$ must be smaller than the relaxation time of the currents. The current is inactive to any fluctuating potentials once the steady correlation is built between the level and leads (see the solid circle in Fig.~\ref{fig:evolution}, which coincides with the $I-t$ curve for $V_g=0$). We find that, to control the NESS, one should drive the system before it reaches the steady state but not after. Second, the correlation time of $V_g$ must be finite. As $\tau\to 0$, i.e., $V_g$ becomes a white noise, the stationary current is totally ignorant to the fluctuating potential~(see the empty circle in Fig.~\ref{fig:evolution}), which can be understood as follows. The evolution operator for a given path of $V_g(t)$ from $t_i$ to $t_f$ can be factorized into a series of unitary operators,
\begin{equation}
 \hat U [V_g(t)] = \prod_{j=1}^n \hat U (t_j,t_{j-1}),
\end{equation}
where $t_0=t_i$, $t_n=t_f$, and $t_j-t_{j-1}=\Delta t$. If $V_g(t)$ is a white noise, the values of $V_g(t)$ at different times are independent to each other and then the average of the evolution operator over different paths of $V_g(t)$, defined as $\hat{\bar U}= \int D[V_g(t)] p[V_g(t)] \hat U [V_g(t)]$, is found to be
\begin{equation}\label{whitenoiseevolution}
 \hat{\bar U} = \int \prod_{j=1}^n d V_g(t_j) p (V_g(t_j)) \hat U (t_j,t_{j-1}),
\end{equation}
where $p[V_g(t)]$ denotes the probability of a path. As $\Delta t\to 0$, we have $\hat U(t_j,t_{j-1}) \approx 1 -i \hat H(t_j) \Delta t$. Then Eq.~(\ref{whitenoiseevolution}) gives $\hat{\bar U}= \hat U[V_g(t)=0]$, indicating that the evolution of the system in the presence of a white noise is exactly the same as that without any noise.

\begin{figure}
\includegraphics[width=1.0\linewidth]{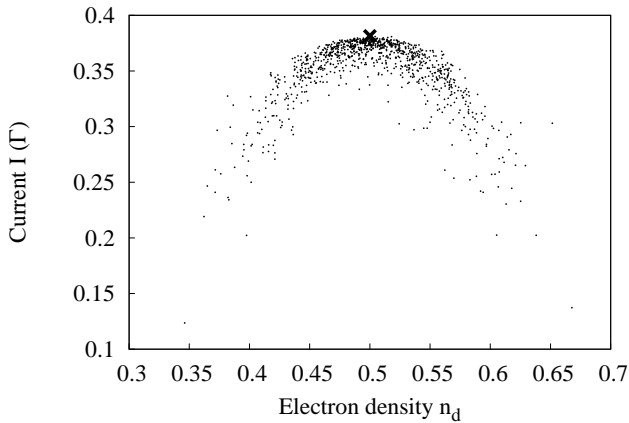}
\caption{A scatter plot of the fluctuations of the stationary current $I$ and the electron density $n_d$ around their means. The voltage bias is set to $V=5\Gamma$. The fluctuating potential $V_g(t)$ is applied during $[0, 1/\Gamma]$ with the correlation time being $\tau=0.002/\Gamma$ and the strength being $R=5\Gamma$. The cross ``$\times$'' represents the stationary current and electron density at $V_g(t)\equiv 0$.}\label{fig:scatterplot}
\end{figure}
The stationary current in the presence of a fluctuating potential satisfying the above two conditions depends upon the history of $V_g(t)$. We then estimate the statistics of the stationary current by simulating $V_g(t)$ for the given $R$ and $\tau$ for many times. In Fig.~\ref{fig:scatterplot}, we show the scatter plot of the fluctuations of both the current and the electron density $n_d= \lim_{t\to\infty} \langle \hat c^\dag_0(t) \hat c_0(t)\rangle $ at the level around their means. It is obvious that the stationary current distributes in a wide range with an upper limit equal to the current without fluctuating potentials. The current is substantially related to the electron density, indicated by the bell shape of the cloud. The more the electron density $n_d$ deviates from $0.5$, the less the current is. The suppressed current in the presence of a fluctuating potential is then attributed to the temporary deviation of the level from the resonant position, which causes the off-resonance effect accompanied by $n_d$ deviating from $0.5$.

The non-equilibrium stationary current depends on the history of the fluctuating potential with a finite correlation time applied during the transient period, i.e., the tunneling current is multistable. The maximal current is obtained when there is no fluctuating potential or the fluctuating potential is a white noise. It is worth of mentioning that the multistability of the tunneling current found here should be distinguished from the non-uniqueness of the NESS in the dc transport~\cite{dhar}, resulting from that the levels of the bound states are outside the conduction bands of the leads. Since the central level in our model is at the resonant position, and the fluctuation of the level is always inside the conduction bands.

\begin{figure}
\includegraphics[width=1.0\linewidth]{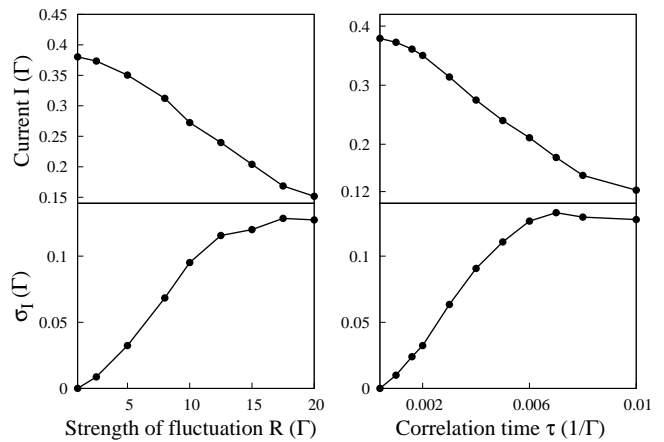}
\caption{[Left panel] The averaged current $I$ and the standard deviation of current $\sigma_I$ as a function of the strength of fluctuation $R$ in $V_g(t)$. The correlation time of $V_g(t)$ is set to $\tau=0.002/\Gamma$. [Right panel] $I$ and $\sigma_I$ as a function of the correlation time $\tau$ of $V_g(t)$. The strength of fluctuation is set to $R=5\Gamma$. In both figures, the voltage bias is $V=5\Gamma$, and the potential $V_g(t)$ is switched on during $[0, 1/\Gamma]$. The solid line connecting the symbols serves as a guide for the eyes.}\label{fig:fluctuationstr}
\end{figure}
{\it Controlling of the non-equilibrium stationary current.}-Finally, we address the effects of the strength of the fluctuation ($R$), the correlation time ($\tau$), and the turn-on ($t_i$) and turn-off time ($t_f$), on the average and standard deviation of currents. By studying the relation between the parameters of the stochastic time series and currents, we find different ways in controlling the non-equilibrium stationary current.

The strength of fluctuation $R$ represents how far the level deviates from the resonant position temporarily, since the variance of $V_g(t)$ is $R^2/(1-\phi^2)$. As $R$ increasing, the central level temporarily moves to a position further away from the resonant point, suppressing the current more due to the off-resonance effect (see Fig.~\ref{fig:fluctuationstr}).

The standard deviation of current, approximately equal to the deviation of the averaged current from its maximum ($V_g(t)=0$), however, increases with $R$. The standard deviation is zero if and only if the current reaches its maximum, i.e., there is no fluctuating potential. The scatter plot of $(I,n_d)$ has the bell shape (see Fig.~\ref{fig:scatterplot}). For small $R$, the deviation of the current from its resonant value is small, so that the points in the scatter plot accumulate around the cross (tested, but not shown in the figure). Then the averaged current is close to its maximum and its standard deviation is small. As $R$ increasing, the points in the scatter plot diffuse into two wings of the cloud, then the standard deviation increases, while the averaged current is reduced.

\begin{figure}
\includegraphics[width=1.0\linewidth]{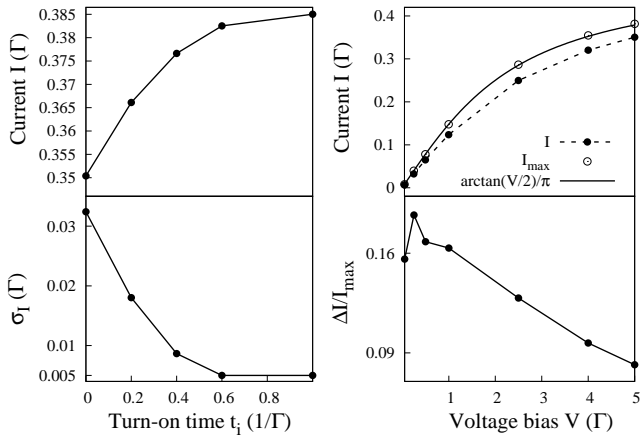}
\caption{[Left panel] The averaged current and the standard deviation of current as a function of the turn-on time $t_i$ of $V_g(t)$. The voltage bias is set to $V=5\Gamma$. The strength, the correlation time, and the lasting time of $V_g(t)$ are set to $R=5\Gamma$, $\tau=0.002/\Gamma$, and $t_f-t_i=1/\Gamma$, respectively. [Right panel] The top right is the current as a function of the voltage bias in the presence of a fluctuating potential (the solid circle, labeled by $I$) with $\tau=0.002/\Gamma$ and $R=5\Gamma$ switched on during the period $[0, 1/\Gamma]$, compared with that without fluctuating potentials (the empty circle, labeled by $I_{max}$). The latter coincides well with the function $I=(\Gamma/\pi) \arctan [V/(2\Gamma)]$ in the wide band limit (the solid line). The bottom right shows the percent of the current reduction, i.e., the ratio of $\Delta I=I_{max}-I$ to $I_{max}$. The other line connecting the symbols serves as a guide for the eyes.}\label{fig:turnontime}
\end{figure}
The correlation time $\tau$ represents how fast the fluctuating potential oscillates. In the white noise limit, the potential oscillates too fast so that it cannot be felt by the electrons moving in it, and then the current does not reduce as we analyze above. Increasing $\tau$ will increase the life of a temporary level that deviates from the resonant position and then suppress the current by the off-resonance effect. The averaged current should then decrease as the correlation time $\tau$ of $V_g(t)$ increasing, as we see in Fig.~\ref{fig:fluctuationstr}. Increasing either $R$ or $\tau$ will strongly suppress the averaged current. At $R=20\Gamma$ for $\tau=0.002/\Gamma$ or $\tau=0.01/\Gamma$ for $R=5\Gamma$, the current reduces only to be 30\% of its maximum. A fluctuating potential can thus be a candidate of suppressing the tunneling current.

Our results then provide a possible explanation on the reduced current in the transport through quantum dots due to the temporary deviation of the level from the resonant position by a fluctuating potential. The suppression of the Kondo resonance~\cite{rosch} provides another example, in which the fluctuating potential comes from the fluctuations of the charges in the dot exerting forces to the electrons moving through it.

Finally we clarify the effects of the turn-on time $t_i$ on the current. In addition to the first condition for the suppressed current, that is, the fluctuating potential must be applied before the system relaxes to its steady state, we find that the most efficient way of changing the non-equilibrium stationary current is to apply the fluctuating potential at the time that the correlation between leads and the level begins to build. Increasing the turn-on time will gradually raises the averaged current to its maximum, at the same time decreasing the standard deviation of current (see Fig.~\ref{fig:turnontime}).

The averaged current as a function of voltage bias is studied and compared with that of no fluctuating potentials in Fig.~\ref{fig:turnontime}. In the absence of fluctuating potentials, the $I-V$ curve coincides well with the function $I=(\Gamma/\pi) \arctan [V/(2\Gamma)]$ in the wide band limit. A fluctuating potential significantly suppresses the current, but keeps the shape of the $I-V$ curve invariant. The percent of the current reduction keeps finite in the range of the voltage bias (see the bottom right panel of Fig.~\ref{fig:turnontime}), showing that the fluctuating potential suppresses the current both in the linear response regime and beyond it.

{\it Experimental proposal.}-We now show that the quantum dot device made of the metal surface electrodes on a heterostructure~\cite{heinzel} is a potential candidate for observing the predicted suppression of the tunneling current when the dot is subject to a randomly fluctuating potential. Our model describes qualitatively the transport through a quantum dot tuned into the Coulomb blockade regime, where the level spacing in the dot is much larger than the level width $\Gamma$ such that only a single level is used to shuttle the electrons. In experiments, the fluctuating potential $V_g$ can be controlled by the gate voltage $\tilde{ \mathcal{ V}}_g$, satisfying $V_g = \alpha \tilde{ \mathcal{ V}}_g$ with $\alpha$ measured in the experiment~\cite{goldhaber-gordon}. The gate voltage should first be adjusted to sufficiently negative so that the quantum dot pinches off. Then the gate voltage is tuned to the resonant value corresponding to a conductance peak. At the same time, a sequence of pulses generated according to $\tilde{\mathcal{V}}_g(t+\Delta t)=\phi \tilde{\mathcal{V}}_g(t) + \tilde R W_t $ is applied to the gate voltage with the parameters $\phi=exp(-\Delta t/\tau)$ and $\tilde R= R/\alpha$. And the pulse length $\Delta t$ should be much smaller than $1/\Gamma$, where the level width $\Gamma$ can be determined by measuring the width of the conductance peak. The current in the presence of short pulses can be measured by using the techniques in excited-state spectroscopy~\cite{hanson}. 
The measured tunneling current as a function of the parameters $R$ and $\tau$ is expected to qualitatively coincide with Fig.~\ref{fig:fluctuationstr}.

{\it Conclusions.}-
In summary, we have investigated the effect of the manually generated fluctuating potential simulated by the autoregressive model on the tunneling current. Our results predict that the fluctuating potential with a finite correlation time, when applied before the non-equilibrium steady state is built up, efficiently suppresses the stationary current through a resonant level. The suppression of the current is attributed to the off-resonance effect due to the correlation between the stationary current and the electron density at the level. The off-resonance effect is caused by the temporary deviation of the level from the resonant position, and happens both in the linear response regime and beyond it. After the fluctuating potential is withdrawn and the level returns to the resonant position, the stationary current will not recover its resonant value. Our results show that the stationary current heavily depends on the history of the fluctuating potential during the transient period, and thus provide an efficient way of controlling the NESS without changing the temperatures and chemical potentials of the reservoirs.

We acknowledge the useful discussions with Guy Cohen. Gao X. was supported by the NSF of China under Grant No. 11174253 and by the Zhejiang Provincial Natural Science Foundation under Grant No. R6110175. S. Xu was supported by the NSF of China under Grant No. 71103161.

\appendix
\section*{Supplementary material}

\section*{Method}

We employ the numerical excitation operator method~\cite{wang13} to calculate the tunneling current $I(t)$ and its stationary value $I=\lim_{t\to \infty} I(t)$. The numerical method is described as follows.

The current is expressed as
\begin{equation}
 I(t) = -g_c \left( \textbf{Im} \langle \hat c^\dag_{-1}(t) \hat c_0(t) \rangle + \textbf{Im} \langle \hat c^\dag_{0}(t) \hat c_1 (t) \rangle \right),
\end{equation}
where the brackets denote the expectation value to the initial state. At the initial time $t=0$, the two leads are in their own equilibrium states respectively and they are both decoupled to the level which is empty. To obtain $I(t)$, we first calculate the field operators $\hat c^\dag_{-1}(t), \hat c_0(t)$ and $\hat c^\dag_{1}(t)$ in the Heisenberg picture and then their expectation values.

\begin{figure}
\includegraphics[width=1.0\linewidth]{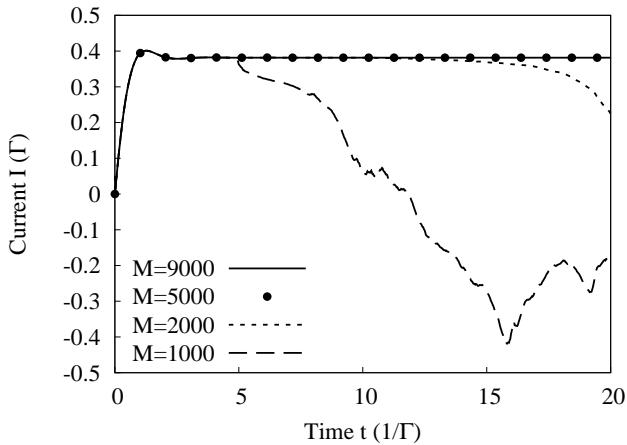}
\caption{The currents calculated by the numerical excitation operator method at different $M$. The potential is $V_g(t)=0$ and the voltage bias is $V=5\Gamma$. Taking $M=5000$ is enough for obtaining the high precision stationary current, since increasing $M$ further to $9000$ will not significantly change the result. The relative difference of the currents at $M=5000$ and $M=9000$ is found to be less than $10^{-5}$.}\label{fig:errorcheck}
\end{figure}
The field operators satisfy the Heisenberg equation $\frac{d}{dt} \hat c^\dag_j(t) = i[\hat H, \hat c^\dag_j(t)] $. Its solution is supposed to be
\begin{equation}
 \hat c^\dag_j (t) = \sum_k W_{jk}(t) \hat c^\dag_k.
\end{equation}
Because $\hat c^\dag_j (t+\Delta t)=e^{i\hat H \Delta t} \hat c^\dag_j(t) e^{-i\hat H \Delta t} $, we obtain
\begin{equation}\label{iterativerelation}
\begin{split}
\sum_k W_{jk}(t+\Delta t) & \hat c^\dag_k \\
= \sum_k W_{jk}(t) & \bigg( \hat c^\dag_k+ i\Delta t [\hat H, \hat c^\dag_k] + \frac{(i\Delta t)^2}{2} [\hat H,[\hat H,\hat c^\dag_k]] \\ & + O(\Delta t^3) \bigg).
\end{split}
\end{equation}
The small time step $\Delta t$ is taken, and thus, the terms $O(\Delta t^3)$ can be neglected. $W_{jk}(t+\Delta t)$ can also be expressed as a linear function of $W_{jk}(t)$:
\begin{equation}
\begin{split}
 W_{jk}(t+\Delta t)= & W_{jk}(t) +i\Delta t \sum_l W_{jl}(t) G_{lk} \\ & -\frac{\Delta t^2}{2} \sum_{l,m} W_{jl}(t) G_{lm} G_{mk}  ,
\end{split}
\end{equation}
where the coefficients $G_{kl}$ are defined by the commutators $[\hat H, \hat c^\dag_k]= \sum_l G_{kl}\hat c^\dag_l $. Then the propagators $W_{jk}(t)$ at an arbitrary time are worked out by an iterative algorithm starting from $t=0$ when $W_{jk}(0)= \delta_{j,k}$ and moving forward $\Delta t$ at each step. The error caused by a finite $\Delta t$ is of the order $O(\Delta t^3)$, and then can be made negligible by setting $\Delta t$ very small. One can also keep the higher order terms of $\Delta t$ in Eq.~(\ref{iterativerelation}). Keeping the terms in order of $O(\Delta t^2)$ has been proved to be efficient for obtaining the stationary current.

At each step, the non-zero propagators $W_{jk}(t)$ are stored and used to calculate the propagators at next step. The number of non-zero propagators will increase quickly. A truncation scheme is then applied so that only a fixed number of non-zero propagators (the number is denoted by $M$) with the largest magnitudes are kept. This truncation scheme is critical for obtaining the stationary current within a reasonable computation time. The error caused by a finite $M$ can be made negligible by setting $M$ large enough. To obtain the current at the longer time, a larger $M$ should be chosen. The value of $M$ depends on the relaxation time of the current. In this letter, setting $M$ to several thousands is enough for obtaining the high precision stationary current (see Fig.~\ref{fig:errorcheck}).

\end{document}